\input{aipcheck}

\documentclass[
    ,final            
  ]
  {aipproc}

\layoutstyle{6x9}

\begin{document}

\title{$0^+$ states in the large boson number limit of the Interacting Boson Approximation model}

\classification{21.10.Re, 21.60.Fw, 05.70.Fh}
\keywords      {Interacting Boson Approximation model, geometric collective model, 
shape/phase transitions, order parameter}

\author{Dennis Bonatsos}{
  address={Institute of Nuclear Physics, National Centre for Scientific Research
   ``Demokritos'', GR-15310 Aghia Paraskevi, Attiki, Greece}
}

\author{E. A. McCutchan}{
  address={Physics Division, Argonne National Laboratory, Argonne, Illinois 60439, USA}
}

\author{R. F. Casten}{
  address={Wright Nuclear Structure Laboratory, Yale University, New Haven, Connecticut 06520-8124, USA} }

\begin{abstract}

Studies of the Interacting Boson Approximation (IBA) model for large boson numbers have been triggered 
by the discovery of shape/phase transitions between different limiting symmetries of the model. 
These transitions become sharper in the large boson number limit, revealing previously unnoticed regularities, which also survive to a large extent for finite boson numbers, corresponding to 
valence nucleon pairs in collective nuclei. It is shown that energies of $0_n^+$ states grow linearly with their ordinal number $n$ in all three limiting symmetries of IBA [U(5), SU(3), and O(6)]. Furthermore, it is proved that the narrow transition region separating the symmetry triangle of the IBA into a spherical and a deformed region is described quite well by the degeneracies $E(0_2^+)=E(6_1^+)$, $E(0_3^+)=E(10_1^+)$, $E(0_4^+)=E(14_1^+)$, while the energy ratio $E(6_1^+) /E(0_2^+)$ turns out to be a simple,
empirical, easy-to-measure effective order parameter, distinguishing between first- and second-order
transitions. The energies of $0_n^+$ states near the point of the first order shape/phase transition 
between U(5) and SU(3) are shown to grow as n(n+3), in agreement with the rule dictated by the relevant critical point symmetries resulting in the framework of special solutions of the Bohr Hamiltonian. The underlying partial dynamical symmetries and quasi-dynamical symmetries are also discussed.   

\end{abstract}

\maketitle



The Interacting Boson Approximation (IBA) model \cite{IA}, describing collective phenomena in atomic nuclei
in terms of $s$ and $d$ bosons (of angular momentum 0 and 2, respectively), 
 is known to possess an overall U(6) symmetry, 
containing three different dynamical symmetries, U(5), SU(3), and O(6), corresponding to 
near-spherical (vibrational), axially symmetric prolate deformed (rotational), and 
soft with respect to axial asymmetry ($\gamma$-unstable) nuclei respectively. These limiting 
symmetries are shown at the vertices of the symmetry triangle \cite{Casten} of the model, shown in Fig.~1. 
   
\begin{figure} 
\includegraphics[height=60mm]{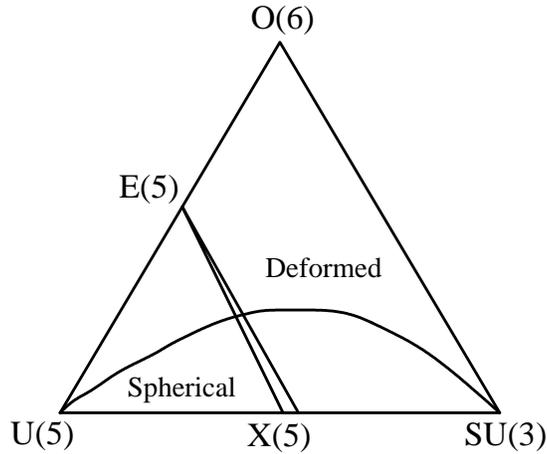}
\caption{IBA symmetry triangle with the three dynamical symmetries. The critical point models E(5) and X(5) are placed
close to the phase transition region (slanted
lines). The solid curve indicates the Alhassid-Whelan arc of regularity. Adopted from Ref. \cite{PRL}.}
\end{figure}
   
An energy functional can be obtained \cite{IZC} in the classical limit of the model, through the use of 
the coherent state formalism \cite{GK,DSI}. Studying this energy functional in the framework of catastrophe theory one can see \cite{Feng} that a first order phase transition (in the Ehrenfest classification) is predicted to occur 
between the limiting symmetries U(5) and SU(3), while a second order phase transition is expected 
between U(5) and O(6). We refer to these transitions as shape/phase transitions.   
A narrow shape coexistence region is then predicted \cite{IZC} in the symmetry triangle of the IBA, 
separating the spherical and deformed phases. The shape coexistence region shrinks into the 
point of second order phase transition as the U(5)-O(6) line is approached, as shown in Fig.~1. 

Shape/phase transitions have been considered recently in the framework of the geometric collective 
model \cite{BM} as well, in which the collective variables $\beta$ and $\gamma$ are used. 
The second order transition between U(5) and O(6) has been described by the E(5) critical point symmetry
\cite{IacE5}, using in the Bohr Hamiltonian \cite{Bohr} a potential independent of $\gamma$ 
and having the shape of an infinite 
square well potential in $\beta$, while the first order transition between U(5) and SU(3) has been described by the X(5) critical point symmetry \cite{IacX5}, 
using in the Bohr Hamiltonian a potential of the form $u(\beta)+v(\gamma)$,
with $u(\beta)$ having again the shape of an infinite square well potential in $\beta$, and $v(\gamma)$ 
being a steep harmonic oscillator centered around $\gamma=0$. E(5) and X(5) are shown in Fig.~1, close to the 
points of the second and first order phase transitions of the IBA, respectively. 

The IBA model predictions have also been studied using different measures of chaotic behavior.
It has been found that chaotic behavior prevails over most of the symmetry triangle, with the 
noticeable exception of two regions, in which highly regular behavior is observed \cite{AW}. One of them is 
located along the U(5)-O(6) line, and its existence is due to the O(5) symmetry known to survive along 
the whole line \cite{Talmi}. The other regular region is connecting U(5) and SU(3) through a narrow path 
within the triangle, called the Alhassid--Whelan arc of regularity \cite{AW} (also shown in Fig.~1), 
the symmetry implying its existence being yet unknown. 

\begin{table}
\caption{Irreducible representations (irreps) of SU(3) and O(6) and
the corresponding energy of the excited $0^+$ bandheads. 
$N$ stands for the boson number, $N_B$. Adopted from Ref. \cite{PRL}.}
\begin{tabular}{cccc|cc}
\hline
\multicolumn{4}{c}{SU(3)} & \multicolumn{2}{c}{O(6)}\\
\hline

Irrep ($\lambda$,$\mu$) & $E$($0^+$) & Irrep ($\lambda$,$\mu$) & $E$($0^+$) &
Irrep ($\sigma$) & $E$($0^+$) \\

\hline

(2$N$,0)    & 0                  &             &                    & ($N$)   & 0         \\
(2$N$-4,2)  & 1                  &             &                    & ($N$-2) & 1         \\
(2$N$-8,4)  & (4$N$-6)/(2$N$-1)  & (2$N$-6,0)  & (4$N$-3)/(2$N$-1)  & ($N$-4) & 2         \\
(2$N$-12,6)  & (6$N$-15)/(2$N$-1)  & (2$N$-10,2) & (6$N$-10)/(2$N$-1) & ($N$-6) & 3-(3/$N$) \\
(2$N$-16,8) & (8$N$-28)/(2$N$-1) & (2$N$-14,4) & (8$N$-21)/(2$N$-1) & ($N$-8) & 4-(8/$N$) \\

\hline
\end{tabular}
\end{table}

States with zero angular momentum are of particular interest, since centrifugal effects are 
absent, facilitating the detection of underlying symmetries. In the U(5) limit of the IBA, 
the energies of $0^+$ states increase linearly with the number of $d$ bosons, corresponding to their
phonon number. In the SU(3) limit, the energies of $0^+$ bandheads are determined by the eigenvalues 
of the second order Casimir invariant of SU(3), the results shown in Table~1. In the O(6) limit, 
they are determined by the second order Casimir invariant of O(6), the results also shown in Table~1.
From Table~1 it is clear that in the limit of large boson numbers $N_B$, the energies of $0^+$ 
bandheads in all three dynamical symmetries of the IBA grow linearly, $E= An$. 
  
\begin{figure} 
\includegraphics[height=90mm]{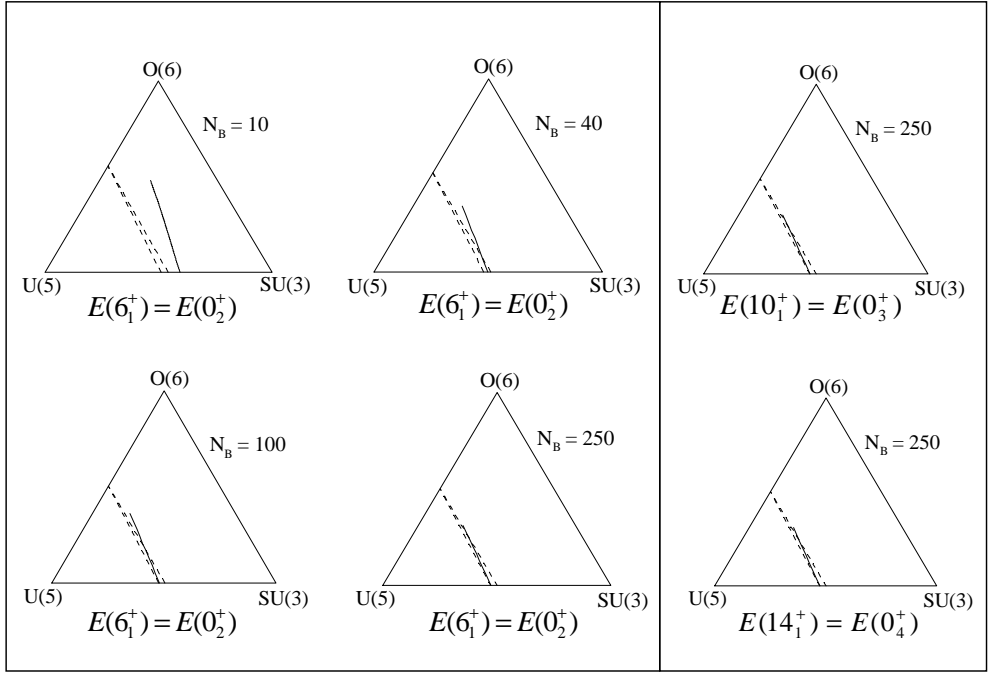}
\caption{ (Left) Line of degeneracy between the
$0_2^+$ and $6_1^+$ levels for $N_B$ = 10, 40, 100,
and 250 in the IBA triangle. (Right) Line of degeneracy between
the $0_3^+$ and $10_1^+$ levels for $N_B$ = 250
(top) and between the $0_4^+$ and $14_1^+$ levels 
 for $N_B$ = 250 (bottom) in the IBA triangle. The dashed
lines denote the critical region in the IBA obtained in the large
$N_B$ limit from the intrinsic state formalism. Adopted from Ref. \cite{PRL100}.}
\end{figure}

IBA numerical calculations have been performed using the recently developed code
IBAR \cite{IBAR,Liz}, which can handle large boson numbers. The standard two-parameter IBA Hamiltonian 
\cite{Werner} has been used, depending on the parameters $\zeta$ and $\chi$. Examining 
the degeneracy $E(0_2^+)=E(6_1^+)$, 
a hallmark of the X(5) critical point symmetry, we find \cite{PRL100} that its locus in the IBA triangle 
is a straight line, falling withing the coexistence region for large boson numbers, as seen in Fig.~2. 
Similar results are obtained for the degeneracies $E(0_3^+)=E(10_1^+)$ and $E(0_4^+)=E(14_1^+)$,
also shown in Fig.~2.  

\begin{figure} 
\includegraphics[height=60mm]{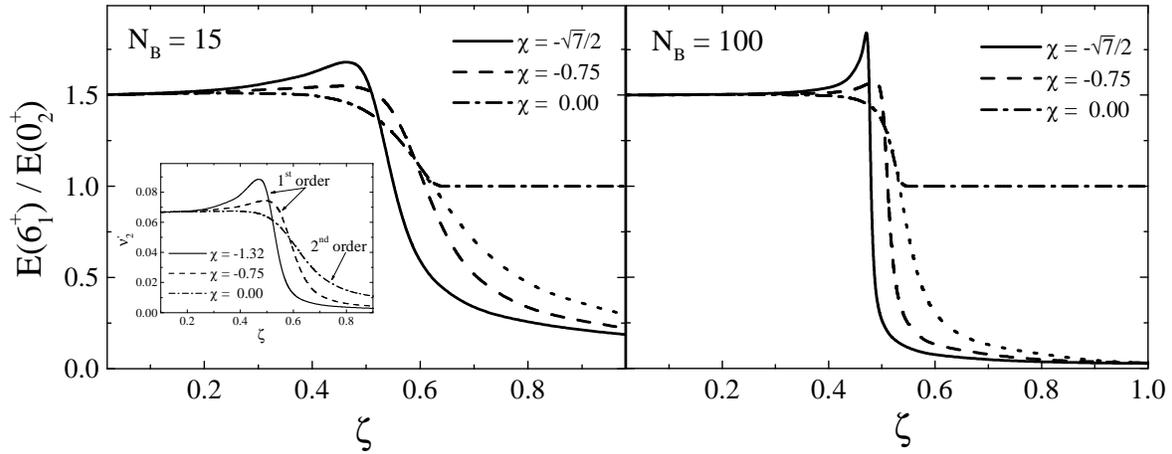}
\caption{ The ratio $E$($6_1^+$)/$E$($0_2^+$) as a
function of $\zeta$ for three values of $\chi$ for (a) $N_B$ = 15
and (b) $N_B$ = 100. The inset to (a) shows the corresponding
behavior for the order parameter $\nu_2^{\prime}$~\cite{IZ}. Adopted from Ref. \cite{PRL100}.}
\end{figure}

\begin{figure} 
\includegraphics[height=50mm]{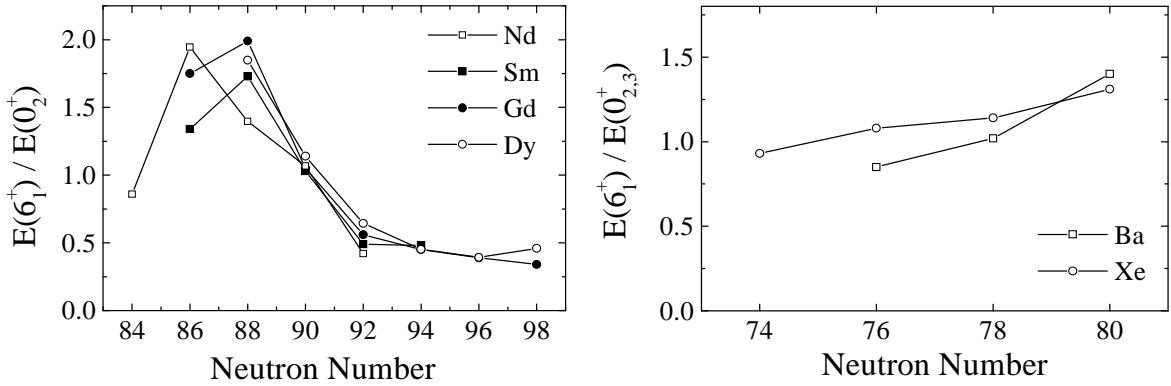} 
\caption{(a) Experimental $E$($6_1^+$)/$E$($0_2^+$)
ratio as a function of neutron number for the Nd, Sm, Gd, and Dy
isotopes. (b) Same for the Xe and Ba isotopes. For smaller neutron
numbers, the $0_3^+$ state was taken in the ratio if its $B$($E$2)
decay was consistent with the $\sigma$ = $N-2$ state. This
corresponds to $N$ = 74 in Xe and $N$ = 76,78 in Ba. Valence (hole) neutron number 
increases to the left. Adopted from Ref. \cite{PRL100}.}
\end{figure}

Plotting the ratio $E(6_1^+)/E(0_2^+)$ \cite{PRL100}, related to the first of the degeneracies 
mentioned above, we see in Fig.~3 that it exhibits the behavior expected \cite{IZ} for an order 
parameter of a first (second) order phase transition for $\chi=-1.32$ ($\chi=0$). It is remarkable that experimental data 
around the $N=90$ isotones, which are known to be the best empirical examples of X(5) 
\cite{CZX5,Kruecken,Tonev,Dewald,review},
shown in Fig.~4, exhibits a behavior very similar to the one expected for a first order phase transition, while 
data around $^{134}$Ba, the best example of E(5) \cite{CZE5,review}, shows the behavior expected for  a
second order transition.  

\begin{table}
\caption{(Left) Energies of $0^+$ states in the E(5), Z(5), and X(5)
models. Energies on the left are in units of $E$($2_1^+$) = 1.0,
while in the column Norm, in units $E$($0_2^+$) = 1.0. The
normalized results are identical for each of the models. The
column IBA-Norm gives the normalized $0^+$ energies for a large
$N_B$ IBA calculation near the critical point. (Middle) Same for the Z(4) model.
(Right) Same for the X(3) model. Adopted from Ref. \cite{PRL}.}
\begin{tabular}{c|c|c|c|c|c||c|c||c|c}
\hline

$0_i^+$ & E(5) & Z(5) & X(5) & Norm & IBA-Norm & Z(4) & Norm & X(3) & Norm \\
\hline
$0_1^+$ & 0 & 0 & 0 & 0  & 0 & 0 & 0 & 0 & 0 \\
$0_2^+$ &  3.03 &  3.91 &  5.65 & 1.0  & 1.0  &  2.95 & 1.0  &  2.87 & 1.0  \\
$0_3^+$ &  7.58 &  9.78 & 14.12 & 2.50 & 2.48 &  7.60 & 2.57 &  7.65 & 2.67 \\
$0_4^+$ & 13.64 & 17.61 & 25.41 & 4.50 & 4.62 & 13.93 & 4.71 & 14.34 & 5.00 \\
$0_5^+$ & 21.22 & 27.39 & 39.53 & 7.00 & 7.13 & 21.95 & 7.43 & 22.95 & 8.00 \\ 
$0_6^+$ & 30.31 & 39.12 & 56.47 &10.00 & 9.85 & 31.65 &10.72 & 33.47 &11.67 \\
\hline
\end{tabular}
\end{table}

\begin{table}
\caption{Order $\nu$, dimension, $D$, of the model space and $\nu$ for $J^{\pi}=0^+$ states in the geometrical models E(5), X(5), Z(5), Z(4), and X(3). $J$ is the spin of the level, $\tau=J/2$, and $n_{w}$ is the wobbling quantum number \cite{BM} which is zero for $0^+$ states.}
\renewcommand{\arraystretch}{2.0}
\begin{tabular}{lccc|lccc}
\hline

Model & $\nu$ & $\;$$\;$$\;$$\;$ D $\;$$\;$$\;$$\;$ & $\nu_{J=0^+}$ &
Model & $\nu$ & $\;$$\;$$\;$$\;$ D $\;$$\;$$\;$$\;$ & $\nu_{J=0^+}$ \\

\hline

E(5) & $\tau$ + $\frac{3}{2}$ & 5 & $\frac{3}{2}$ &   &   &   &   \\

\hline

X(5) & $\sqrt{\frac{J(J+1)}{3} + \frac{9}{4}}$ & 5 &  $\frac{3}{2}$ &
Z(5) & $\frac{\sqrt{J(J+4)+3n_w(2J-n_w)+9}}{2}$ & 5 &  $\frac{3}{2}$ \\

\hline 

X(3) & $\sqrt{\frac{J(J+1)}{3}+\frac{1}{4}}$ & 3 & $\frac{1}{2}$ & 
Z(4) & $\frac{\sqrt{J(J+4)+3n_w(2J-n_w)+4}}{2}$ & 4 &  1 \\
\hline

\end{tabular}
\end{table}

Now we turn attention to energies of $0^+$ states within the critical point symmetries
E(5) and X(5), mentioned above, as well as within Z(5) \cite{Z5}, a similar solution of the Bohr 
Hamiltonian, also using an infinite square well potential in $\beta$, with $\gamma \approx \pi/6$. 
As seen in Table~2, although the relevant energies 
look different in each model when normalized to the energy of the $2_1^+$ state, they become 
identical if normalized to the energy of $0_2^+$ \cite{PRL}. Moreover, they increase as $n(n+3)$.
Similarly, energies of $0^+$ states within Z(4) \cite{Z4} [similar to Z(5), but with $\gamma$ fixed 
to $\pi/6$] increase as $n(n+2.5)$. Within X(3) \cite{X3} [similar to X(5), but with $\gamma$ fixed to 0],
they increase as $n(n+2)$, as also shown in Table~2. These results can be easily interpreted \cite{PRL}  
by taking into account the order of the Bessel functions appearing as eigenfunctions in these models,
given in Table~3, as well as the fact that the spectra of the Bessel functions $J_\nu$ are found to increase 
approximately as $n(n+\nu+3/2)$, this result being exact only for $\nu=1/2$ \cite{PRL}. 
Consideration of the second order Casimir operator of E($n$) \cite{Barut,Z4}, 
the Euclidean algebra in $n$ dimensions,
shows \cite{PRL} that the present situation is a partial dynamical symmetry \cite{AL}
of Type I \cite{Leviatan}, in which some of the states
(the $0^+$ states in the present case) preserve all the relevant symmetry.   
The recent conjecture \cite{Macek} of a partial SU(3) dynamical symmetry underlying the Alhassid--Whelan 
arc of regularity is also receiving attention.  

It is a nontrivial result \cite{PRL} that the IBA near the critical point of the first order transition also 
yields energies of $0^+$ states increasing as $n(n+3)$, i.e. in the same way as critical point 
symmetries based on infinite square well potentials in 5 dimensions (degrees of freedom) predict,
as shown in Table~2.  

In conclusion, $0^+$ states in the large boson number limit of the IBA exhibit many interesting 
properties, as well as degeneracies to non-zero angular momentum states, which invite further investigations 
into determining of the symmetries underlying these regularities.

\begin{theacknowledgments}
  
  Work supported in part U.S. DOE Grant No.
DE-FG02-91ER-40609 and  under Contract DE-AC02-06CH11357.
  
\end{theacknowledgments}


\begin{thebibliography}{9}

\bibitem{IA}
F.~Iachello, and A.~Arima, \emph{The Interacting Boson Model}, Cambridge 
University Press, Cambridge, 1987. 

\bibitem{Casten}
R.~F. Casten, \emph{Nuclear Structure from a Simple Perspective},
Oxford University Press, Oxford, 1990.

\bibitem{PRL}
D.~Bonatsos, E.~A. McCutchan, and R.~F. Casten, \emph{Phys. Rev. Lett.}, accepted.  

\bibitem{IZC}
F.~Iachello, N.~V. Zamfir, and R.~F. Casten, \emph{Phys. Rev. Lett.} \textbf{81}, 1191--1194 (1998).

\bibitem{GK}
J.~N. Ginocchio, and M.~W. Kirson, \emph{Phys. Rev. Lett.} \textbf{44}, 1744--1747 (1980)

\bibitem{DSI}
A.~E.~L. Dieperink, O.~Scholten, and F.~Iachello, \emph{Phys. Rev. Lett.} \textbf{44}, 1747--1750 (1980). 

\bibitem{Feng}
D.~H. Feng, R.~Gilmore, and S.~R. Deans, \emph{Phys. Rev. C} \textbf{23}, 1254--1258 (1981). 

\bibitem{BM}
A.~Bohr, and B.~R. Mottelson, \emph{Nuclear Structure, Vol. II}, Benjamin, New York, 1975. 

\bibitem{IacE5} 
F.~Iachello, \emph{Phys. Rev. Lett.}  \textbf{85}, 3580--3583 (2000). 

\bibitem{Bohr} 
A.~Bohr, \emph{Mat. Fys. Medd. K. Dan. Vidensk. Selsk.} \textbf{26}, no. 14 (1952). 

\bibitem{IacX5} 
F.~Iachello, \emph{Phys. Rev. Lett.} \textbf{87},  052502 (2001). 

\bibitem{AW} 
Y.~Alhassid, and N.~Whelan, \emph{Phys. Rev. Lett.} \textbf{67}, 816--819 (1991).

\bibitem{Talmi} 
A.~Leviatan, A.~Novoselsky, and I.~Talmi, \emph{Phys. Lett. B} \textbf{172}, 144--148 (1986). 

\bibitem{IBAR}
R.~J. Casperson, IBAR code (unpublished). 

\bibitem{Liz}
E.~Williams, R.~J. Casperson, and V.~Werner, \emph{Phys. Rev. C} \textbf{77}, 061302(R) (2008). 

\bibitem{Werner}
V.~Werner, N.~Pietralla, P.~von Brentano, R.~F. Casten, and R.~V. Jolos, \emph{Phys. Rev. C}
\textbf{61}, 021301(R) (2000).

\bibitem{PRL100}
D.~Bonatsos, E.~A. McCutchan, R.~F. Casten, and R.~J. Casperson, \emph{Phys. Rev. Lett.} \textbf{100},
142501 (2008).

\bibitem{IZ} 
F.~Iachello, and N.~V. Zamfir, \emph{Phys. Rev.Lett.} \textbf{92}, 212501 (2004).

\bibitem{CZX5} 
R.~F. Casten, and N.~V. Zamfir, \emph{Phys. Rev. Lett.} \textbf{87}, 052503 (2001). 

\bibitem{Kruecken} 
R.~Kr\"ucken, {\it et al.}, \emph{Phys. Rev. Lett.} \textbf{88}, 232501 (2002).

\bibitem{Tonev} 
D.~Tonev, {\it et al.}, \emph{Phys. Rev. C} \textbf{69}, 034334 (2004). 

\bibitem{Dewald} 
A.~Dewald, {\it et al.}, \emph{Eur. Phys. J. A} \textbf{20}, 173--178 (2004). 

\bibitem{review}
R.~F. Casten, and E.~A. McCutchan, \emph{J. Phys. G: Nucl. Part. Phys.} \textbf{34}, R285--R320 (2007). 

\bibitem{CZE5}
R.~F. Casten, and N.~V. Zamfir, \emph{Phys. Rev. Lett.} \textbf{85}, 3584--3586 (2000). 

\bibitem{Z5}
D.~Bonatsos, D.~Lenis, D.~Petrellis, and P.~A. Terziev, \emph{Phys. Lett. B} \textbf{588}, 172--179 (2004). 

\bibitem{Z4}
D.~Bonatsos, D.~Lenis, D.~Petrellis, P.~A. Terziev, and I.~Yigitoglu, 
\emph{Phys. Lett. B} \textbf{621}, 102--108 (2005). 

\bibitem{X3}
D.~Bonatsos, D.~Lenis, D.~Petrellis, P.~A. Terziev, and I.~Yigitoglu, 
\emph{Phys. Lett. B} \textbf{632}, 238--242 (2006).  

\bibitem{Barut}
A. O. Barut, and R. Raczka, \emph{Theory of Group Representations and 
Applications}, World Scientific, Singapore, 1986. 

\bibitem{AL}
Y.~Alhassid, and A.~Leviatan, \emph{J. Phys. A: Math. Gen.} \textbf{25},
L1265--L1271 (1992).

\bibitem{Leviatan}
A.~Leviatan, \emph{Phys. Rev. Lett.} \textbf{98}, 242502 (2007).

\bibitem{Macek} 
M.~Macek, P.~Str\'ansk\'y, P.~Cejnar, S.~Heinze, J.~Jolie, and J.~Dobe\v s, 
\emph{Phys. Rev. C} \textbf{75}, 064318 (2007).

\end{thebibliography}
\end{document}